\documentclass[twocolumn,aps,prd,floatfix,%
preprintnumbers,nofootinbib]{revtex4-2}

\usepackage{epsfig}
\usepackage{amsmath,bm}
\newcommand{\I}{{\rm i}}

\begin{document}

\title{N-mode coherence in collective neutrino oscillations}

\author{Georg G.~Raffelt}
\affiliation{Max-Planck-Institut f\"ur Physik
(Werner-Heisenberg-Institut), F\"ohringer Ring 6, 80805 M\"unchen,
Germany}

\date{13 September 2021}

\begin{abstract}
We study two-flavor neutrino oscillations in a homogeneous and
isotropic ensemble under the influence of neutrino-neutrino
interactions. For any density there exist forms of collective
oscillations that show self-maintained coherence. They can be
classified by a number $N$ of linearly independent functions that
describe all neutrino modes as linear superpositions. What is more,
the dynamics is equivalent to another ensemble with the same
effective density, consisting of $N$ modes with discrete energies
$E_i$ with $i=1,\ldots, N$. We use this equivalence to derive the
analytic solution for two-mode (bimodal) coherence, relevant for
spectral-split formation in supernova neutrinos.
\vskip6pt
\noindent In this post-publication version, Eqs.~(B4) and (B5) corrected.
\end{abstract}

\preprint{\hbox to\textwidth{PRD {\bf 83}, 105022 (2011)  \hfill MPP-2011-17}}


\maketitle

\section{Introduction}                               \label{sec:intro}

In the early universe and in collapsing stars, the density of
neutrinos is so large that they produce significant refractive
effects for each other. In a seminal paper~\cite{Pantaleone:1992eq},
Pantaleone showed that this is not just a correction to the usual
matter effect~\cite{Wolfenstein:1977ue, Mikheev:1986gs, Kuo:1989qe},
but causes qualitatively new phenomena. They originate from flavor
off-diagonal refraction caused by neutrino oscillations which
therefore feed back on themselves. One consequence is self-induced
flavor conversion even for a very small mixing angle, caused by an
instability of the interacting ensemble. When the density decreases
slowly, for example in a supernova as a function of radius, this
instability leads to flavor swaps between neutrino spectra in sharp
energy intervals, causing conspicuous ``spectral splits.'' The body
of literature on collective oscillations has immensely grown, so we
cite only some theoretical key papers~\cite{Sigl:1992fn,
Pantaleone:1998xi, Samuel:1993uw, Kostelecky:1993dm,
Kostelecky:1995dt, Samuel:1996ri, Pastor:2001iu, Wong:2002fa,
Friedland:2003dv, Friedland:2006ke, Duan:2005cp, Duan:2006an,
Duan:2007mv, Raffelt:2007yz, Sawyer:2008zs, Hannestad:2006nj,
Balantekin:2006tg, Raffelt:2007cb, Dasgupta:2009mg, Fogli:2007bk,
Fogli:2008pt, Raffelt:2010za}, whereas we refer to a recent
review~\cite{Duan:2010bg} for the torrent of activities in the area
of supernova neutrinos.

One key feature of collective oscillations is that, given suitable
initial conditions, the entire ensemble evolves in a highly
correlated way (self-maintained coherence). We make this concept
more precise, if only for the simplest case of a homogeneous and
isotropic gas with two mixed flavors and ignoring ordinary matter.
Moreover, we mostly consider fixed density, although we use
motivations related to adiabatic changes of density.

\begin{figure}[b]
\includegraphics[width=0.40\columnwidth]{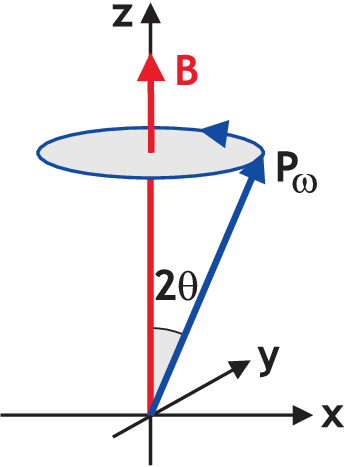}
\caption{Precession of a polarization vector ${\bf P}_\omega$ with
frequency $\omega$
around the $z$--direction (the mass direction in flavor space). The
initial orientation is in the weak-interaction direction, tilted by
twice the mixing angle relative to the
$z$--direction.\label{fig:precession}}
\end{figure}

In the absence of neutrino-neutrino interactions, every mode evolves
independently of the others. The ensemble may be described by
two-spinors in flavor space $\psi_E(t)$, occupation-number matrices
$\varrho_{E}(t)$, or the corresponding polarization vectors ${\bf
P}_E(t)$---effectively there are three important numbers for every
mode: the amplitudes of the two flavor components and their relative
phase. We always use polarization vectors and label the modes with
their vacuum oscillation frequency $\omega=\Delta m^2/2E$, where
negative $\omega$ denote antineutrinos. Vacuum oscillations are
described by the polarization vector of a given mode precessing with
frequency $\omega$ around a direction given by a unit vector ${\bf
B}$, the mass direction in flavor space. If neutrinos begin in
weak-interaction eigenstates, all ${\bf P}_\omega$ are initially
tilted relative to ${\bf B}$ by twice the vacuum mixing angle
(Fig.~\ref{fig:precession}). Each ${\bf P}_\omega$ precesses with a
different frequency and so eventually the modes distribute
themselves uniformly on the precession cone centered on ${\bf B}$,
approaching complete kinematical decoherence.

Neutrino-neutrino interactions provide an additional force in terms
of the global ${\bf P}=\int d\omega\,{\bf P}_\omega$, leading to the
equations of motion (EoMs)
\begin{equation}\label{eq:precession}
\dot{\bf P}_\omega=(\omega {\bf B}+\mu{\bf P})\times{\bf P}_\omega\,.
\end{equation}
Here $\mu\sim\sqrt2\,G_{\rm F}n_\nu$ is a typical interaction
energy, its exact value depending on the normalization of the
polarization vectors. In the limit $\mu\to\infty$, and if initially
all ${\bf P}_\omega$ are aligned in the flavor direction, they
``stick together'' and precess around ${\bf B}$ with the common
frequency $\langle\omega\rangle$. These ``synchronized
oscillations'' are the simplest case of self-maintained coherence
\cite{Kostelecky:1995dt, Pastor:2001iu, Wong:2002fa}.

Starting with such a configuration, $\mu$ can slowly decrease as in
the expanding universe or with distance from a supernova. The
ensemble continues to oscillate coherently, but the ${\bf P}_\omega$
slowly separate and fan out in a plane that precesses around ${\bf
B}$. Such a pure precession mode can exist for any strength of $\mu$
\cite{Duan:2007mv, Raffelt:2007cb}. If $\mu$ decreases adiabatically
to zero, the final configuration is that all ${\bf P}_\omega$ with
$\omega$ above some frequency $\omega_{\rm split}$ point in the
positive ${\bf B}$ direction, the others in the opposite direction,
forming a ``spectral split.''

\begin{figure}[ht]
\includegraphics[width=0.75\columnwidth]{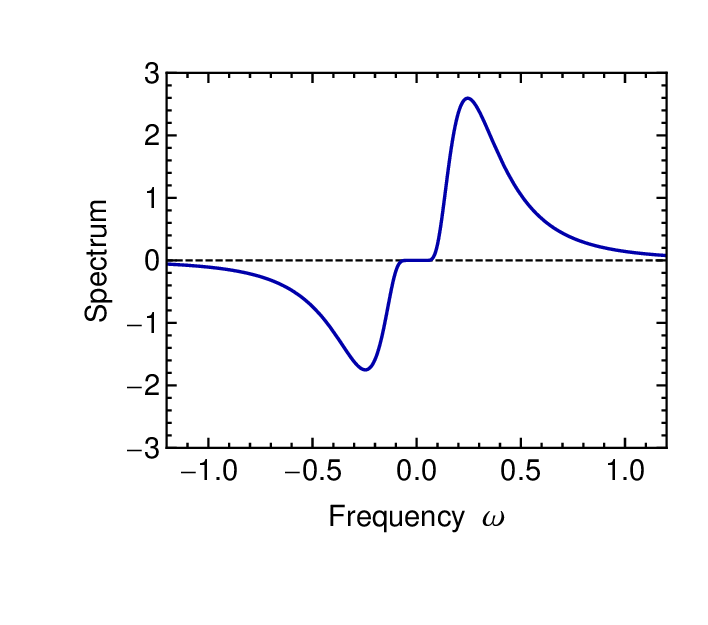}
\caption{Neutrino spectrum ($z$-component of ${\bf P}_\omega$)
in the variable $\omega=T/E$, assuming that initially only one
flavor is populated with a Fermi-Dirac distribution
(degeneracy parameter
$\eta=0.2$). The integral over positive $\omega$ is normalized to unity.
\label{fig:spectrum0}}
\end{figure}

More complicated forms of coherent motion occur in the context of
multiple split formation~\cite{Dasgupta:2009mg}. A generic example
is a gas of neutrinos and antineutrinos of a given flavor with a
thermal distribution. Figure~\ref{fig:spectrum0} shows a Fermi-Dirac
distribution as a function of the re-scaled frequency $\omega=T/E$.
In this way $\omega$ is a dimensionless variable of order unity. The
neutrino flux streaming from a supernova core contains an excess of
$\nu_e$ over $\bar\nu_e$ because the trapped electron-lepton number
needs to be carried away. We mimic this situation schematically by
assuming a small degeneracy parameter $\eta=0.2$, corresponding to
$n_{\bar\nu}\approx0.70\,n_{\nu}$. The region around $\omega=0$
corresponds to the high-$E$ tail and the high-$\omega$ tails
(small~$E$) fall off as~$\omega^{-4}$. The mixing angle is taken to
be very small and so the mass and flavor directions are almost
identical. Notice that ${\bf P}_\omega$ pointing up means one
flavor, pointing down the other. In the flavor-isospin convention
\cite{Duan:2006an, Dasgupta:2009mg}, the interpretation is reversed
for antineutrinos, explaining that the spectrum in
Fig.~\ref{fig:spectrum0} is negative for $\omega<0$.

\begin{figure}
\includegraphics[width=0.79\columnwidth]{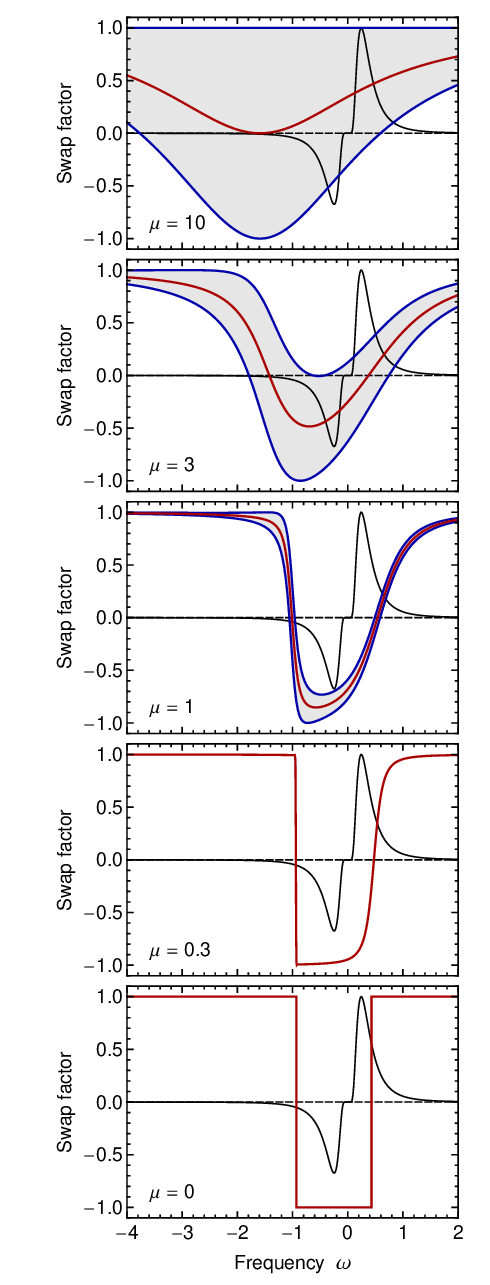}
\vskip-3pt
\caption{Maximum, minimum and average swap factor for
the thermal spectrum of Fig.~\ref{fig:spectrum0} if the interaction
strength begins with $\mu=10$ and then decreases adiabatically to 0.
Snapshots for $\mu=10$, 3, 1, 0.3 and 0 (top to bottom).
The initial spectrum is overlaid using a vertically
compressed scale relative to Fig.~\ref{fig:spectrum0}.\label{fig:swapexample1}}
\end{figure}

Without $\nu$-$\nu$ interactions only small-amplitude vacuum
oscillations occur, whereas for non-zero $\mu$ the system is
unstable. The evolution is nicely illustrated by showing the
instantaneous ``swap factor'' $z_\omega(t)$, i.e.\ the
$\omega$-dependent factor by which the original spectrum must be
multiplied to obtain the instantaneous $z$--component of ${\bf
P}_\omega(t)$, i.e.\ $P_\omega^z(t)=z_\omega(t)P_\omega^z(0)$
\cite{Dasgupta:2009mg} . The entire spectrum oscillates coherently.
In the top panel of Fig.~\ref{fig:swapexample1} we show numerical
results ($\mu=10$) for the maximum and minimum $z_\omega$ during an
oscillation period as well as their average.

In this example and in any unstable spectrum, $z_\omega(t)$ is
initially a Lorentzian of a certain width $\lambda$ that oscillates
like an inverted gravitational pendulum with natural frequency equal
to the same $\lambda$~\cite{Dasgupta:2009mg}. When $\mu$ decreases
adiabatically, $z_\omega$ changes, the oscillation amplitude
decreases, and at $\mu=0$ takes on square-well shape: a spectral
swap with two sharp splits at its edges has formed. Several
snapshots are shown in Fig.~\ref{fig:swapexample1}.

These examples show ``coherence'' in some intuitive sense, but what
precisely does this mean? We here define coherent motion as a form
of evolution where neighboring modes do not separate from each
other. Kinematical decoherence means that different ${\bf
P}_\omega(t)$ within a frequency interval $\delta\omega$ eventually
separate by a large amount, no matter how small $\delta\omega$. If
we coarse-grain the ensemble with a bin width $\delta\omega$, the
coarse-grained $\langle {\bf P}_\omega\rangle$ becomes shorter than
${\bf P}_\omega$ as time goes on~\cite{Raffelt:2010za}. On the other
hand, coherence means that the coarse-grained ensemble remains
identical with the original one, no matter how much time has passed.
Pure precession is a trivial example because the polarization
vectors remain static relative to each other, whereas in the example
of Fig.~\ref{fig:swapexample1}, distant ${\bf P}_\omega(t)$ move
wildly relative to each other, yet neighboring ones do not separate
as time goes~on.

Independent evolution of neighboring modes, no matter how close
their frequencies, implies that every ${\bf P}_\omega(t)$ is
linearly independent. For vacuum oscillations this is obvious
because each ${\bf P}_\omega(t)$ is a harmonic function of frequency
$\omega$. On the other hand, kinematical decoherence can not occur
if every ${\bf P}_\omega(t)$ is a linear combination of a small
number $N$ of independent functions. In this case the system
performs what we call \hbox{$N$-mode} coherent oscillations. The
infinitely dimensional function space required for kinematical
decoherence has reduced to a small number of dimensions.

What is more, the same function space is spanned by the solutions
${\bf J}_i(t)$ of another system ($i=1,\ldots,N$) with vacuum
frequencies $\omega_i$ and the same effective density. So we
consider the EoMs of a new system
\begin{equation}\label{eq:EoMcarrier}
\dot {\bf J}_i=
(\omega_i{\bf B}+\mu{\bf J})\times{\bf J}_i\,.
\end{equation}
The functions ${\bf J}_i(t)$ can be found such that
\begin{equation}\label{eq:selfconsistency}
{\bf P}(t)={\bf J}(t)\,,
\end{equation}
meaning that both systems are dynamically equivalent. The original
modes ${\bf P}_\omega(t)$ and the linearly independent ``carrier
modes'' ${\bf J}_i(t)$ span the same function space: each ${\bf
P}_\omega(t)$ is a certain linear combination of the ${\bf J}_i(t)$.
Notice that usually the carrier modes can not be interpreted as a
coarse-grained representation of the original ensemble, but rather
are more abstract.

The idea that coherence in collective neutrino oscillations
corresponds to linear dependence among different modes and that the
evolution is dynamically equivalent to a small ensemble of carrier
modes provides a simple picture for self-maintained coherence. Our
primary interest is to use this approach for the construction of
explicit classes of solutions, notably for bimodal coherence.

We begin our study in Sec.~\ref{sec:nmode} with the equations of
motion, the idea of $N$-mode coherence, and the concept of carrier
modes. Then we proceed to the explicit cases of single-mode,
two-mode and multi-mode coherence in
Secs.~\ref{sec:singlemode}--\ref{sec:multimode} and conclude in
Sec.~\ref{sec:conclusions}.

\section{N-Mode Coherence}                           \label{sec:nmode}

\subsection{Equations of motion}

The two-flavor evolution of a homogeneous and iso\-tro\-pic neutrino
gas can be described by flavor polarization vectors ${\bf
P}_\omega(t)$, where $\omega=\Delta m^2/2E$ is the vacuum
oscillation frequency. ${\bf P}_\omega$ pointing up means one
flavor, pointing down the other. In the flavor-isospin convention
\cite{Duan:2006an}, the interpretation is reversed for antineutrinos
($\omega<0$). The flavor evolution under the influence of vacuum
oscillations and neutrino-neutrino interactions is given by
Eq.~(\ref{eq:precession}). This precession equation has been stated
many times~\cite{Sigl:1992fn, Pantaleone:1998xi, Kostelecky:1995dt,
Pastor:2001iu, Wong:2002fa, Raffelt:2007yz, Raffelt:2007cb,
Duan:2010bg}, most recently in great detail in
Ref.~\cite{Raffelt:2010za}, and we use it here without further ado.

A few points still deserve explicit mention. The polarization
vectors derive from an ensemble average and thus are classical
quantities. Our EoMs rely on a mean-field assumption: a given
neutrino is supposed to feel only the average effect of all others,
ignoring fluctuations or quantum correlations. (The quantum to
classical transition was studied in Refs.~\cite{Friedland:2003dv,
Friedland:2006ke}.) The length of ${\bf P}_\omega(t)$ is not fixed
by a quantization condition, but chosen at convenience. The value of
$\mu$ depends on this choice.

The EoMs conserve energy and derive from the classical Hamiltonian
\begin{equation}\label{eq:classHam}
H=\int \omega{\bf B}\cdot{\bf P}_\omega\,d\omega
+\frac{\mu}{2}\,{\bf P}^2\,.
\end{equation}
Here, $H$ and all ${\bf P}_\omega$ are functions of the canonical
coordinates and momenta. The evolution of any function $F$ on phase
space is given by the Poisson bracket $\dot F=[F,H]$. It is crucial
that the ${\bf P}_\omega$ play the role of classical spins, or
rather isospins. They obey the angular-momentum Poisson bracket
$[P_{\omega,x},P_{\omega',y}] =\delta(\omega-\omega')\,P_{\omega,z}$
and cyclic permutations. Notice that $[\,{\cdot}\,,\,{\cdot}\,]$ is
not a commutator---classical variables commute in the sense
$P_{x}P_{y}=P_{y}P_{x}$. The overall ${\bf P}$ is the total angular
momentum (in flavor space). Collective oscillations in the
mean-field approximation are thus equivalent to a set of interacting
classical spins. However, the simplicity of our Hamiltonian is
deceiving because it encapsulates all the complications of
collective flavor oscillations.

\subsection{Rotating frames}

\label{sec:rotatingframes}

Another conserved quantity is ${\bf P}$ projected on ${\bf B}$,
equivalent to angular-momentum conservation in the symmetry
direction. Therefore, we always work in the mass basis and use
coordinates where the $z$--direction coincides with ${\bf B}$
(Fig.~\ref{fig:precession}). Usually all ${\bf P}_\omega$ begin in a
weak-interaction eigenstate and thus in a common direction that is
tilted relative to ${\bf B}$ by twice the mixing angle.

The conservation of $P_z={\bf B}\cdot{\bf P}$ implies that the
internal motion of the ensemble, up to overall precession, does not
depend on $P_z$. Depending on convenience, we can study our system
from the perspective of a frame co-rotating with some
frequency~\cite{Duan:2005cp}. We may choose $\omega_{\rm c}=\mu P_z$
and without loss of generality study the equivalent EoMs
\begin{equation}\label{eq:EoM-transverse}
\dot{\bf P}_\omega=(\omega {\bf B}+\mu{\bf P}_\perp)\times{\bf P}_\omega\,,
\end{equation}
where ${\bf P}_\perp$ is the part of ${\bf P}$ transverse to ${\bf
B}$. We will frequently encounter the degeneracy between a shift of
the neutrino $\omega$ spectrum by some amount $\omega_{\rm c}$ and
$\mu P_z$. On the Hamiltonian level, such shifts amount to
subtracting the conserved quantity $\omega_{\rm c}{\bf B}\cdot{\bf
P}$.

Shifting a spectrum such as Fig.~\ref{fig:spectrum0} by an amount
$\omega_{\rm c}$ modifies the interpretation because
negative-frequency modes (anti-neutrinos) become positive-frequency
modes or vice versa. A mode that used to be occupied by $\bar\nu_e$
is then interpreted as being occupied, for example, by $\nu_\mu$.
However, this modified interpretation does not affect the abstract
dynamics, and we can always shift back at the very end. The
possibility to shift anti-neutrino and neutrino modes seamlessly
into each other is the main advantage of the flavor isospin
convention.

\subsection{Linear dependence of different modes}

Collective neutrino oscillations consist of strong correlations
among all modes and we have advanced that this means that the ${\bf
P}_\omega(t)$ depend linearly on a small number $N$ of independent
functions. To be more precise, we use $N$ to denote the number of
nontrivial functions because in addition the constant ${\bf B}$ will
be seen to appear, so ${\bf P}_\omega(t)$ actually depends on $N+1$
functions.

It is illuminating to diagnose $N$ in numerical examples. The
ensemble is represented by a large discrete set ${\bf P}_i(t)$ with
$i=1,\ldots,n$ and $n\gg 1$. To find the number of linearly
independent functions on a time interval \hbox{$t_1<t<t_2$} we
calculate the Gram matrix \hbox{$G_{ij}=\int_{t_1}^{t_2}dt\,{\bf
P}_{i}(t)\cdot{\bf P}_{j}(t)$}. Its rank reveals the number of
linearly independent functions. In reality, one finds $N+1$ large
eigenvalues of $G_{ij}$ and the rest very much smaller. They can be
made ever smaller by increasing the degree of adiabaticity, i.e.\
slowing down $\mu(t)$ and decreasing the mixing angle, i.e.\ making
the initial polarization vectors more collinear with ${\bf B}$.
Within constraints of numerical accuracy, the chosen time interval
is irrelevant to find $N+1$ if the ${\bf P}_i(t)$ are analytic
functions, but of course influences the numerical realization of the
Gram matrix.

We have always found the expected~$N$. The formation of a single
spectral swap as in Fig.~\ref{fig:swapexample1} corresponds to
bimodal coherence at any stage, i.e.\ the Gram matrix was found to
have rank 3. The evolution from a double-crossed spectrum towards
two swaps~\cite{Dasgupta:2009mg} corresponds to four-mode coherence
(rank~5), and so forth.

\subsection{Carrier modes}

If every ${\bf P}_\omega(t)$ is a linear combination of $N$
independent functions and the constant ${\bf B}$ we may choose any
set ${\bf P}_{\omega_i}(t)$ with $i=1,\ldots,N$ as a basis if they
are not accidentally degenerate. We then have ${\bf P}(t)=\int
d\omega\,{\bf P}_\omega(t)=a_0{\bf B}+\sum_{i=1}^N\,a_i {\bf
P}_{\omega_i}(t)$ as a unique linear combination. We define ${\bf
J}_i(t)\equiv a_i {\bf P}_{\omega_i}(t)$, obeying $\dot{\bf
J}_i=(\omega_i{\bf B}+a_0\mu{\bf B}+\mu{\bf J})\times{\bf J}_i$
where ${\bf J}=\sum_{i=1}^N{\bf J}_i$. In a rotating frame we can
transform away $a_0\mu{\bf B}$, or rather, we may choose another set
${\bf P}_{\omega_i'}(t)$ where $a_0'=0$.

So we can always find a set of linearly independent functions ${\bf
J}_i(t)$ with $i=1,\ldots,N$ that obey Eq.~(\ref{eq:EoMcarrier}),
fulfill the matching condition Eq.~(\ref{eq:selfconsistency}), and
together with ${\bf B}$ span the same function space as the original
ensemble ${\bf P}_\omega(t)$. The internal dynamics of our system is
the same if we study the co-rotating EoMs of
Eq.~(\ref{eq:EoM-transverse}). In this case the matching condition
is ${\bf P}_\perp(t)={\bf J}_\perp(t)$ and the complication about
the $a_0$ coefficient disappears.

We can express each ${\bf P}_\omega(t)$ of the original ensemble as
a linear combination of the chosen ``carrier modes'' ${\bf J}_i(t)$
\begin{equation}\label{eq:linearcombi}
{\bf P}_\omega(t)=b_0{\bf B}+\sum_{i=1}^N b_i\,{\bf J}_i(t)\,.
\end{equation}
This function must obey the original EoM of
Eq.~(\ref{eq:precession}). Inserting Eq.~(\ref{eq:linearcombi}) on
the l.h.s.\ of Eq.~(\ref{eq:precession}) and using
Eq.~(\ref{eq:EoMcarrier}) yields $\sum_{i=1}^N b_i(\omega_i{\bf
B}+\mu{\bf J})\times {\bf J_i}$. On the r.h.s.\ of
Eq.~(\ref{eq:precession}) we use ${\bf P}={\bf J}$ and
Eq.~(\ref{eq:linearcombi}) such that $-b_0\mu{\bf B}\times\mu{\bf
J}+(\omega{\bf B}+\mu{\bf J})\times\sum_{i=1}^N b_i{\bf J}_i$. With
${\bf J}=\sum_{i=1}^N {\bf J}_i$ in the first term and after
collecting everything we find
\begin{equation}
0={\bf B}\times\sum_{i=1}^N\bigl[
b_i\,(\omega-\omega_i)-b_0\mu\bigr]{\bf J}_i(t)\,.
\end{equation}
The functions ${\bf J}_i(t)$ are linearly independent by
construction and not proportional to ${\bf B}$. Therefore, we find
the coefficient $b_i=b_0\mu/(\omega-\omega_i)$ and thus
\begin{equation}\label{eq:linearcombi2}
{\bf P}_\omega(t)=b_0\left[
{\bf B}+\sum_{i=1}^N \frac{\mu}{\omega-\omega_i}\,{\bf J}_i(t)\right]\,.
\end{equation}
The required linear combination is unique up to an overall
proportionality factor. (This linear combination is an example of a
more general transformation briefly discussed in
Appendix~\ref{sec:transformedvectors}.)

The expression, Eq.~(\ref{eq:linearcombi2}), is singular at the
carrier frequencies. However, it is only the direction that matters,
so we introduce the transformed carrier spectrum
\begin{equation}\label{eq:Jhilbert2}
\bar{\bf J}_\omega(t)=\left[{\bf B}+
\sum_{i=1}^N\frac{\mu}{\omega-\omega_i}\,{\bf J}_i(t)\right]
\prod_{i=1}^N\,(\omega-\omega_i)\,,
\end{equation}
where the overbar has nothing to do with antiparticles. Introducing
the spectrum $g_\omega$ with $|{\bf P}_\omega|=|g_\omega|$, the
original ensemble in terms of the carrier modes is
\begin{equation}
{\bf P}_\omega=g_\omega\,\frac{\bar{\bf J}_\omega}{\bar{J}_\omega}\,,
\end{equation}
where $\bar J_\omega=|\bar{\bf J}_\omega|$.

We usually consider initial conditions such that the ${\bf
P}_\omega$ are almost collinear with ${\bf B}$ and we ask for the
probability for a neutrino mode $\omega$ to stay in its original
flavor, perhaps after an adiabatic change of $\mu$. This information
is contained in the $z$--component of $\bar{\bf J}_\omega$ and we
call
\begin{equation}
z_\omega(t)=\frac{{\bf B}\cdot\bar{\bf J}_\omega(t)}{\bar{J}_\omega}
\end{equation}
the time-dependent ``swap factor.''

Our discussion is motivated by the spectral swaps and splits that
form when the effective density encoded in $\mu$ slowly decreases to
zero. Since we are dealing with a Hamiltonian system, we expect the
adiabatic change of the parameter $\mu$ to deform the solution, but
that it remains coherent if it was initially coherent. For $\mu\to0$
we have $\bar{\bf J}_\omega\propto{\bf B}$, implying that in the end
all polarization vectors are collinear with ${\bf B}$. Conceivably
one could define \hbox{$N$-mode} coherence by the very property of
developing $N$ spectral splits when $\mu$ decreases adiabatically to
zero.

Our main interest is to use these ideas to construct explicit
classes of $N$-mode coherent solutions, notably the most general
bimodal solution. For a set of functions ${\bf J}_i(t)$ fulfilling
Eq.~(\ref{eq:EoMcarrier}), the matching condition
Eq.~(\ref{eq:selfconsistency})~is
\begin{equation}\label{eq:selfconsistency1}
{\bf J}(t)=\int d\omega\,g_\omega\,
\frac{\bar{\bf J}_\omega}{\bar{J}_\omega}\,.
\end{equation}
For a given spectrum $g_\omega$ and interaction strength $\mu$ this
condition restricts possible sets of carrier modes to provide an
$N$-mode coherent solution.

\section{Single-Mode Coherence}                 \label{sec:singlemode}

A first trivial case is what we may call zero-mode coherence, where
all ${\bf P}_\omega$ are exactly aligned with ${\bf B}$ and
therefore stay that way. This configuration also appears as a
limiting case of other forms of oscillation.

The first non-trivial case is single-mode coherence, identical with
pure precession~\cite{Duan:2007mv, Raffelt:2007cb}. The single
carrier mode ${\bf J}(t)$ precesses around ${\bf B}$ with some
co-rotation frequency $\omega_{\rm c}$ and obeys $\dot{\bf
J}=\omega_{\rm c}{\bf B}\times{\bf J}$ where the nonlinear term
$\mu{\bf J}\times{\bf J}=0$ has dropped out. According to
Eq.~(\ref{eq:Jhilbert2}) the transformed spectrum is $\bar{\bf
J}_\omega=(\omega-\omega_{\rm c})\,{\bf B}+\mu{\bf J}$. The matching
condition ${\bf P}={\bf J}$ then implies
\begin{equation}
\bar{\bf J}_\omega=(\omega-\omega_{\rm c})\,{\bf B}+\mu{\bf P}
\end{equation}
and thus $\bar{\bf J}^2=(\omega-\omega_{\rm c}+\mu P_z)^2+(\mu
P_\perp)^2$, where ${\bf P}_\perp$ denotes the part of ${\bf P}$
transverse to~${\bf B}$. The matching condition
Eq.~(\ref{eq:selfconsistency1}) is, in agreement with
Ref.~\cite{Raffelt:2007cb},
\begin{eqnarray}\label{eq:precessionconsisteny}
P_z&=&\int d\omega\,g_\omega\,
\frac{\omega-\omega_{\rm c}+\mu P_z}
{\sqrt{(\omega-\omega_{\rm c}+\mu P_z)^2+(\mu P_\perp)^2}}\,,
\nonumber\\*
P_\perp&=&\int d\omega\,g_\omega\,
\frac{\mu P_\perp}
{\sqrt{(\omega-\omega_{\rm c}+\mu P_z)^2+(\mu P_\perp)^2}}\,.
\end{eqnarray}
We have used the fact that all ${\bf P}_\omega$ lie in the plane
spanned by ${\bf B}$ and ${\bf P}$.

We could have studied our system from the perspective of a rotating
frame as discussed in Sec.~\ref{sec:rotatingframes}. In the above
expressions we recognize that nothing changes if we absorb $\mu P_z$
in the definition of $\omega_{\rm c}$.

For a given spectrum $g_\omega$ we can use these equations in
different ways. If $P_z$ is fixed by an initial condition we may ask
for the value of $P_\perp$ and $\omega_{\rm c}$ corresponding to a
specified value of $\mu$. In particular, for $\mu\to 0$ we can find
the split frequency for a given $P_z$. For explicit solutions in
this sense and further discussions we refer to the
literature~\cite{Duan:2007mv, Raffelt:2007cb}.

We may also use these conditions to construct all possible pure
precession solutions for a given $g_\omega$. To this end we
introduce the parameters $\kappa=\mu P_\perp$ and
$\gamma=\omega_{\rm c}-\mu P_z$, leading to the general pure
precession solution in the form of a two-parameter family depending
on $\gamma$ and $\kappa$,
\begin{eqnarray}\label{eq:pureprecessionsolution}
P_{\omega,z}&=&g_\omega\,
\frac{\omega-\gamma}
{\sqrt{(\omega-\gamma)^2+\kappa^2}}\,,
\nonumber\\*
P_{\omega,\perp}&=&g_\omega\,\frac{\kappa}
{\sqrt{(\omega-\gamma)^2+\kappa^2}}\,.
\end{eqnarray}
In addition one needs to specify the precession phase and thus the
instantaneous orientation of the comoving plane. Integrating over
$d\omega$ yields $P_z$ and $P_\perp$ and the corresponding
$\mu=\kappa/P_\perp$ and $\omega_{\rm c}=\gamma+\mu P_z$.

\begin{figure}
\includegraphics[width=0.75\columnwidth]{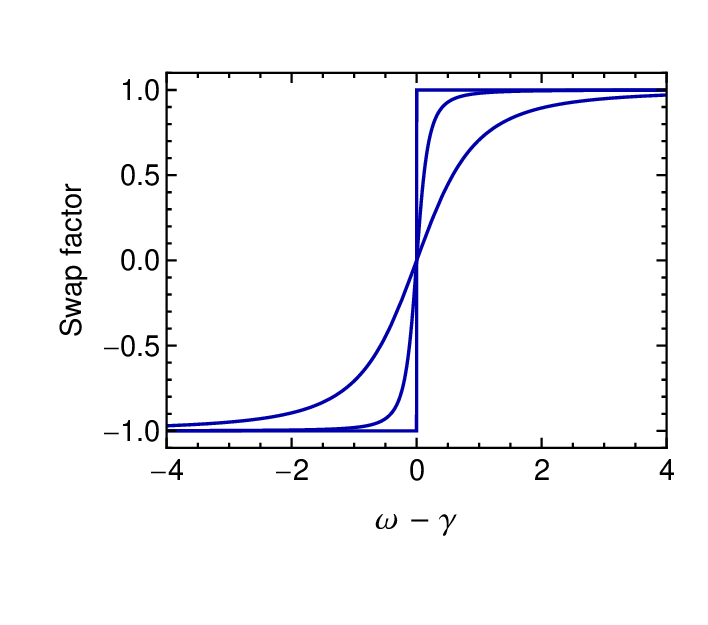}
\caption{Swap factor for pure precession with
$\kappa=0$, 0.5 and~1 according to
Eq.~(\ref{eq:swap-precession}).\label{fig:swap-precession}}
\end{figure}

In terms of the parameters $\kappa$ and $\gamma$, the swap factor
for pure precession is
\begin{equation}\label{eq:swap-precession}
z_\omega=\frac{\omega-\gamma}
{\sqrt{(\omega-\gamma)^2+\kappa^2}}\,.
\end{equation}
Some examples are shown in Fig.~\ref{fig:swap-precession}. The step
occurs at frequency $\gamma$ and the transition region has width
$\kappa$.

For $\kappa\to0$ the swap factor is identical with the sign function
$z_\omega={\rm sgn}(\omega-\gamma)$. In this case the first relation
in Eq.~(\ref{eq:precessionconsisteny}) has the form $P_z=\int
d\omega\,g_\omega\,{\rm sgn}(\omega-\gamma)$ and can be used to find
$\gamma$ for a given $P_z$. The second equation is $ 1=\int
d\omega\,g_{\omega+\gamma}\,\mu/\sqrt{\omega^2+\kappa^2}$ and can be
used to find $\kappa$ explicitly in the limit where $\mu$ is very
small. In this limit, the integral is peaked around $\omega=0$ and
we can approximately pull out $g_\gamma$. The remaining integral
diverges for $\omega\to\pm\infty$, so we cut it off at some
frequencies $\pm\alpha$, leading to $1\sim2\mu g_\gamma
\log(2\alpha/\kappa)$ and implying that for small $\mu$
\begin{equation}\label{eq:kappasteep}
\kappa\propto e^{-1/(2\mu g_{\gamma})}\,.
\end{equation}
In other words, $\kappa$ becomes exponentially small for small $\mu$
or small $g_\gamma$.

\section{Bimodal Coherence}                        \label{sec:bimodal}

\subsection{Matching conditions}

Turning to bimodal coherence, we consider two carrier modes ${\bf
J}_{1,2}(t)$ with frequencies $\omega_{1,2}$ fulfilling the
precession equation, Eq.~(\ref{eq:EoMcarrier}). According to
Eq.~(\ref{eq:Jhilbert2}) the transformed carrier spectrum is
\begin{eqnarray}
\bar{\bf J}_\omega&=&(\omega-\omega_1)(\omega-\omega_2)\,{\bf B}
\nonumber\\
&&{}
+\mu\,(\omega-\omega_2)\,{\bf J}_1+\mu\,(\omega-\omega_1)\,{\bf J}_2\,.
\end{eqnarray}
Two interacting polarization vectors are dynamically equivalent to a
gyroscopic pendulum (Appendix~\ref{sec:twovecs}). To use this
equivalence we write $\omega_1=\omega_{\rm c}-\beta$ and
$\omega_2=\omega_{\rm c}+\beta$, where $\omega_{\rm c}$ is a
suitable co-rotation frequency such that the two carrier modes have
equal but opposite frequencies $\pm\beta$. To simplify notation we
implement $\omega_{\rm c}$ by shifting the spectrum, i.e.\ instead
of $g_\omega$ we use $g_{\omega+\omega_{\rm c}}$ in the integral for
the matching conditions.

From Appendix~\ref{sec:twovecs} we borrow ${\bf Q}={\bf J}_1-{\bf
J}_2+(\beta/\mu)\,{\bf B}$ for the pendulum's radius vector and find
\begin{equation}
\bar{\bf J}_\omega=\omega^2\,{\bf B}
+\mu\,\omega\,{\bf J}+\mu\,\beta\,{\bf Q}\,,
\end{equation}
where ${\bf J}={\bf J}_1+{\bf J}_2$ is the total angular momentum.
To spell out the matching conditions we need a unit vector in the
$\bar{\bf J}_\omega$ direction and thus need
\begin{eqnarray}
\bar{\bf J}_\omega^2&=&\omega^4+(\mu\beta{\bf Q})^2
+2\mu\omega^2(\beta{\bf B}\cdot{\bf Q}+{\textstyle\frac{1}{2}}\,\mu{\bf J}^2)
\nonumber\\
&&{}+2\mu\omega^3{\bf B}\cdot{\bf J}+2\mu^2\beta\omega\,{\bf Q}\cdot{\bf J}\,.
\end{eqnarray}
From Eq.~(\ref{eq:kappadef}) we take the natural frequency
$\lambda^2=\beta\mu|{\bf Q}|$, so the second term is $\lambda^4$.
The bracket in the third term is the energy $E$ of the pendulum. In
the first term on the second line we have ${\bf B}\cdot{\bf J}=J_z$,
a conserved quantity, and in the final term ${\bf Q}\cdot{\bf
J}=|{\bf Q}|\,S$ with $S$ the spin (conserved angular momentum
projection on the pendulum axis). Ordering by powers of $\omega$ we
find
\begin{equation}
\bar{\bf J}_\omega^2=\omega^4+2\mu\,(\omega^3J_z+\omega^2E +\omega\lambda^2 S)
+\lambda^4\,.
\end{equation}
As expected, this quantity is conserved, i.e.\ it does not depend on
the oscillation phase.

We also need the individual components of $\bar{\bf J}_\omega$. We
select directions in the plane transverse to ${\bf B}$ that move
with the system and we call the direction along ${\bf J}_\perp$ the
comoving $y$--direction, where ${\bf J}_\perp$ is the part of ${\bf
J}$ transverse to ${\bf B}$. The comoving $x$--direction is then
orthogonal and hence collinear with ${\bf B}\times{\bf J}$. We
project out the three components by taking the scalar product of
$\bar{\bf J}_\omega$ with ${\bf B}\times{\bf J}$, ${\bf
J}_\perp={\bf J}-J_z{\bf B}$ and ${\bf B}$ and find
\begin{eqnarray}\label{eq:Jxyz}
\bar{J}_{\omega,x}&=&\frac{\mu\beta\,{\bf Q}\cdot({\bf
B}\times{\bf J})}{J_\perp}\,,
\nonumber\\
\bar{J}_{\omega,y}&=&\frac{\mu\omega\,{\bf J}_\perp^2+
\mu\beta\,{\bf Q}\cdot{\bf J}_\perp}{J_\perp}\,,
\nonumber\\
\bar{J}_{\omega,z}&=&\omega^2+\mu\omega J_z+\mu\beta\,{\bf
Q}\cdot{\bf B}\,.
\end{eqnarray}
On the l.h.s.\ of the matching condition
Eq.~(\ref{eq:selfconsistency1}) we have in these comoving
coordinates ${\bf J}=(0,J_\perp,J_z)$ so that
\begin{equation}\label{eq:bimod}
\begin{pmatrix}0\\ 1\\ 0\end{pmatrix}=
\int d\omega\,
\frac{g_{\omega+\omega_{\rm c}}}{\bar{J}_\omega}\,
\begin{pmatrix}1\\ \mu\omega\\ \omega^2\end{pmatrix}\,.
\end{equation}
In the first line we have divided out $\bar J_{\omega,x}$ because it
does not depend on $\omega$, and using this condition then
simplifies the second line, and both are used to find the third. So
we could find the matching conditions without even spelling out the
scalar products in Eq.~(\ref{eq:Jxyz}).

Every solution is given in terms of the four pendulum parameters
$E$, $S$, $J_z$ and $\lambda$, although we have not spelled out
Eq.~(\ref{eq:Jxyz}) in terms of these conserved quantities. In
addition we have the shift frequency $\omega_{\rm c}$ as a fifth
parameter. However, we will see that $\omega_{\rm c}$ and $\mu J_z$
appear only combined, similar to the pure precession case, so
actually there are only four independent parameters. The dynamics is
given in terms of $\cos\theta$, the angle between ${\bf B}$ and
${\bf Q}$, that follows the pendulum differential equation of
Eq.~(\ref{eq:thetadot3}). In addition we can calculate the
precession motion following the usual pendulum treatment
(Appendix~\ref{sec:gyropendulum}).

Given the pendulum parameters and the EoM for $\cos\theta$ we have
all the information needed to describe the motion. However, these
parameters do not allow us to determine the original carrier modes
in a unique way---there are many solutions. Notice, in particular,
that the frequency $\beta$ and the length of ${\bf Q}$ are
degenerate parameters. Two interacting polarization vectors are
equivalent to a pendulum in flavor space, but the reverse mapping is
not unique. In particular, the carrier frequencies are not unique.
In certain limits, however, a particular choice may be most
convenient.

\subsection{Swap parameters}

While the pendulum parameters are well matched to the underlying
dynamics, they are not very intuitive when studying cases such as
the one shown in Fig.~\ref{fig:swapexample1}. To find a more
suitable parametrization we first spell out the swap factor
$z_\omega(t)=\bar{J}_{\omega,z}/\bar J_\omega$ explicitly,
\begin{equation}\label{eq:pendulumswap1}
z_\omega(t)=\frac{\omega^2+\mu\omega J_z+\lambda^2\cos\theta(t)}
{\sqrt{\omega^4+2\mu\,(\omega^3J_z+\omega^2E +\omega\lambda^2 S)
+\lambda^4}}\,.
\end{equation}
Inspecting the numerical results of Fig.~\ref{fig:swapexample1}, the
analytic swap factor for pure precession of
Eq.~(\ref{eq:swap-precession}) motivates the following ansatz for
the average swap factor
\begin{equation}\label{eq:doubleswap1}
\bar z_\omega=
\frac{(\omega-\gamma_1)}{\sqrt{(\omega-\gamma_1)^2+\kappa_1^2}}
\frac{(\omega-\gamma_2)}{\sqrt{(\omega-\gamma_2)^2+\kappa_2^2}}\,,
\end{equation}
where $\gamma_{1,2}$ give the location of the two steps and
$\kappa_{1,2}$ their respective widths.

Comparing the coefficients of the $\omega$ polynomial in the
denominators of Eqs.~(\ref{eq:pendulumswap1})
and~(\ref{eq:doubleswap1}) provides
\begin{eqnarray}\label{eq:mapparameters}
\lambda^4&=&(\gamma_1^2+\kappa_1^2)(\gamma_2^2+\kappa_2^2)\,,
\nonumber\\
\mu J_z&=&-(\gamma_1+\gamma_2)\,,
\nonumber\\
\mu E&=&\frac{(\gamma_1+\gamma_2)^2+2\gamma_1\gamma_2+\kappa_1^2+\kappa_2^2}{2}\,,
\nonumber\\
\mu S&=&-\frac{\gamma_1(\gamma_2^2+\kappa_2^2)+\gamma_2(\gamma_1^2+\kappa_1^2)}{\lambda^2}\,.
\end{eqnarray}
A further simplification is achieved if we express the time
variation of the swap factor in terms of a new dimensionless
variable
\begin{equation}
u(t)=\frac{\lambda^2\,\cos\theta(t)-\gamma_1\gamma_2}{\kappa_1\kappa_2}\,,
\end{equation}
implying
\begin{equation}\label{eq:bimodz}
z_\omega(t)=
\frac{(\omega-\gamma_1)(\omega-\gamma_2)+\kappa_1\kappa_2\,u(t)}
{\sqrt{[(\omega-\gamma_1)^2+\kappa_1^2]
[(\omega-\gamma_2)^2+\kappa_2^2]}}\,.
\end{equation}
Inserting $u(t)$ in the EoM for $\cos\theta(t)$ of
Eq.~(\ref{eq:thetadot3}) and using the parameter mapping of
Eq.~(\ref{eq:mapparameters}) provides
\begin{equation}\label{eq:ueom}
\dot u^2=(1-u^2)\left[(\gamma_1-\gamma_2)^2+(\kappa_1-\kappa_2)^2
+2\kappa_1\kappa_2(1-u)\right]\,,
\end{equation}
where the range of motion is $-1\leq u\leq+1$. Therefore, the
average is $\bar u=0$ and the average swap factor is indeed given by
the proposed expression Eq.~(\ref{eq:doubleswap1}).

The structure of Eq.~(\ref{eq:ueom}) suggests the interpretation
$u=\cos\vartheta$ with $\vartheta(t)$ some abstract angle variable.
In this way $1-u^2=\sin^2\vartheta$ and one finds
\begin{equation}
\dot\vartheta^2=(\gamma_1-\gamma_2)^2+(\kappa_1-\kappa_2)^2
+2\kappa_1\kappa_2(1-\cos\vartheta)\,.
\end{equation}
This angle always advances in the same direction with a modulated
velocity. The EoM is that of a gravitational pendulum with a speed
so large that it moves through the inverted position.

\begin{figure}
\includegraphics[width=0.8\columnwidth]{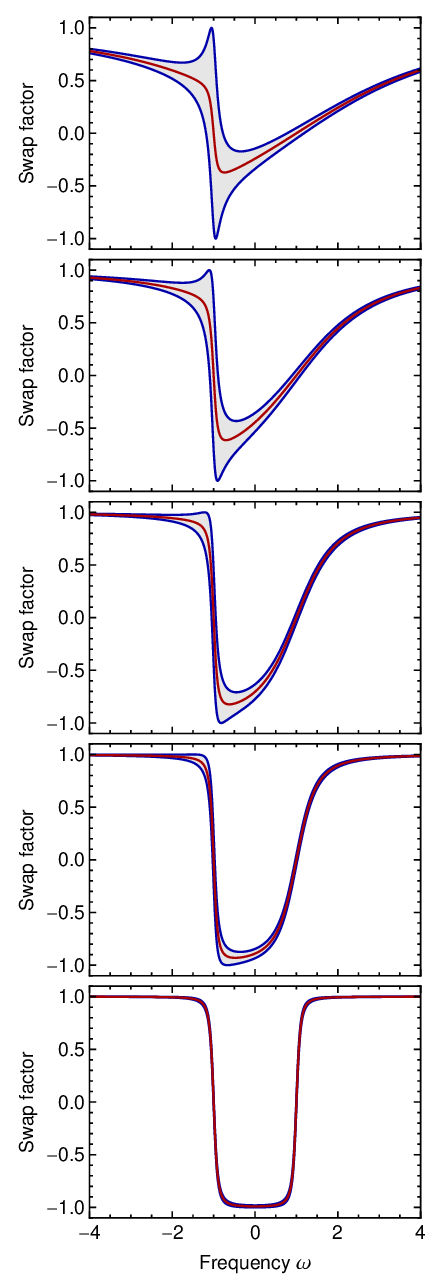}
\caption{Maximum, minimum and average swap factor according to
Eq.~(\ref{eq:bimodz}) using $\gamma_{1}=-1$, $\gamma_{2}=+1$, $\kappa_1=0.1$,
and $\kappa_2=4$, 2, 1, 0.5 and 0.1 (top to bottom).
\label{fig:swapexample2}}
\end{figure}

We show examples for the maximal, minimal and average swap factors,
corresponding to $u=+1$, 0 and $-1$ in Fig.~\ref{fig:swapexample2},
where $\gamma_{1}=-1$, $\gamma_{2}=+1$ and $\kappa_1=0.1$ are held
fixed and we show different cases for $\kappa_2=4$, 2, 1, 0.5 and
0.1 (top to bottom).

We realize that in Eqs.~(\ref{eq:bimodz}) and (\ref{eq:ueom}) the
parameters $\gamma_1$ and $\gamma_2$ always appear either as
$\gamma_1-\gamma_2$ or as $\omega-\gamma_{1,2}$. So we can go to a
frame rotating with some frequency $\omega_{\rm c}$ by shifting
$\omega$ and $\gamma_{1,2}$ by this amount and the solution will be
the same. Since $-(\gamma_1+\gamma_2)/\mu$ has the interpretation of
$J_z$ this means that we can trade $J_z$ for a co-rotation frequency
$\omega_{\rm c}$. Once more this is the same effect discussed in
Sec.~\ref{sec:rotatingframes}. Instead of fixing $J_z=P_z$ it is
enough to match ${\bf P}_\perp={\bf J}_\perp$ and trade the mismatch
of $z$--components for a suitable co-rotation frequency $\omega_{\rm
c}$. Here, in particular, we can adjust $J_z$ such that $\omega_{\rm
c}=0$.

Spelling out the matching conditions in terms of the new parameters,
we find
\begin{equation}\label{eq:bimod2}
\begin{pmatrix}0\\ \mu^{-1}\\ P_z\end{pmatrix}=
\int d\omega\,
\frac{g_{\omega}}{\bar J_\omega}\,
\begin{pmatrix}1\\ \omega\\ (\omega-\gamma_1)(\omega-\gamma_2)\end{pmatrix}\,,
\end{equation}
where
\begin{equation}\label{eq:jbar}
\bar J_\omega=\sqrt{\left[(\omega-\gamma_1)^2+\kappa_1^2\right]
\left[(\omega-\gamma_2)^2+\kappa_2^2\right]}\,.
\end{equation}
For a given spectrum, the first equation establishes a constraint
among the four parameters $\kappa_{1,2}$ and $\gamma_{1,2}$. The
second allows us to calculate the required $\mu$ for this solution,
and the third provides the required $P_z$. We can also use the first
and second condition to simplify the third.

It is also of interest to spell out explicitly all three components
of $\bar{\bf J}_\omega$ of Eq.~(\ref{eq:Jxyz}). For $J_\perp$ we
find
\begin{equation}
(\mu {\bf J}_\perp)^2=
\kappa_1^2+\kappa_2^2-2\kappa_1\kappa_2u\equiv\kappa_u^2
\end{equation}
and
\begin{eqnarray}\label{eq:Jxyz2}
\bar{J}_{\omega,x}&=&\frac{\sqrt{(\gamma_2-\gamma_1)^2+\kappa_u^2}}{\kappa_u}\,
\kappa_1\kappa_2\sqrt{1-u^2}\,,
\nonumber\\
\bar{J}_{\omega,y}&=&\frac{(2\omega-\gamma_1-\gamma_2)\,\kappa_u^2+
(\gamma_2-\gamma_1)(\kappa_2^2-\kappa_1^2)}{2\kappa_u}\,,
\nonumber\\
\bar{J}_{\omega,z}&=&(\omega-\gamma_1)(\omega-\gamma_2)+\kappa_1\kappa_2u\,.
\end{eqnarray}
Divide these components by $\bar J_\omega$ of Eq.~(\ref{eq:jbar}) to
find the unit vector $\bar{\bf J}_\omega/\bar J_\omega$.

For $-\infty<\omega<+\infty$ this vector traces a closed curve on
the unit sphere, where the points $\omega=\pm\infty$ are at the
north pole. This closed curve moves as a function of $\vartheta(t)$.
Notice that for $u=\pm 1$, corresponding to the highest and lowest
swap-factor curves in Fig.~\ref{fig:swapexample1}, the comoving
$x$--component vanishes for all $\omega$, implying that all ${\bf
P}_\omega$ lie in the plane spanned by ${\bf B}$ and ${\bf P}$. The
case $u=-1$ corresponds to the lower swap factor, tracing out all
$z$--values between $-1$ and $+1$, i.e.\ the unit vectors trace out
the great circle spanned by ${\bf B}$ and~${\bf P}$.

\subsection{Pure precession limit: \boldmath$\kappa_1\kappa_2=0$}

\begin{figure}
\includegraphics[width=0.75\columnwidth]{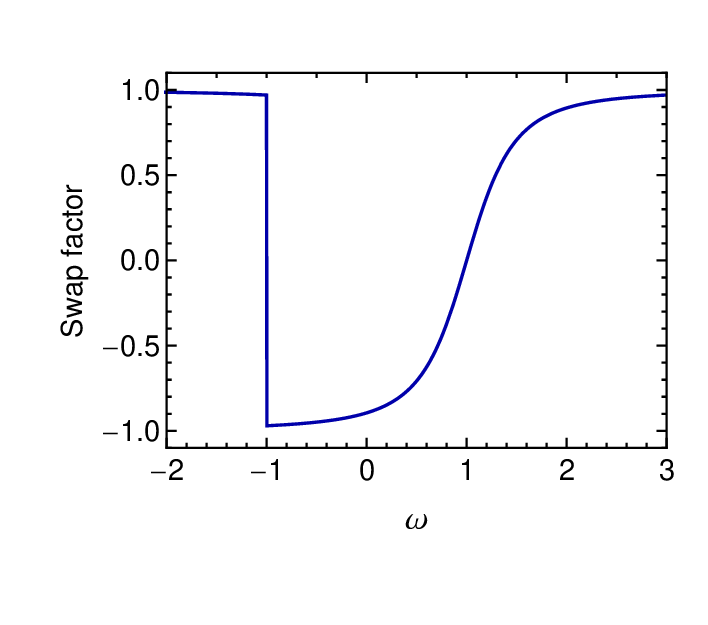}
\caption{Swap factor for pure precession as derived
from the gyroscopic pendulum picture. The chosen parameters are $\gamma_1=-1$,
$\gamma_2=+1$, $\kappa_1=0$ and $\kappa_2=0.5$.\label{fig:gyroprecessionswap}}
\end{figure}

Bimodal solutions are described by two carrier modes, which in turn
are equivalent to a gyroscopic pendulum. One possible form of motion
is pure precession: single-mode coherence appears as a limiting form
of two-mode coherence. In terms of our swap parameters, this case is
described by $\kappa_1=0$ or $\kappa_2=0$ where the solution is
static up to an overall precession. The swap factor, no longer
depending on time, is
\begin{equation}\label{eq:doubleswap3}
z_\omega={\rm sgn}(\omega-\gamma_1)\,
\frac{(\omega-\gamma_2)}
{\sqrt{(\omega-\gamma_2)^2+\kappa_2^2}}
\,,
\end{equation}
where we have used $\kappa_1=0$. In other words, we find the
pure-precession swap factor of Eq.~(\ref{eq:swap-precession}),
augmented with a step function centered on a frequency $\gamma_1$
(Fig.~\ref{fig:gyroprecessionswap}). Therefore, our previous result
is equal to the present one for the choice $\gamma_1\to-\infty$.

\subsection{Pure nutation limit: \boldmath$\kappa_1=\kappa_2$}

Another limiting case is when the swap factor is symmetric around
$\omega_{\rm c}=\frac{1}{2}(\gamma_1+\gamma_2)$, i.e.\ when
$\kappa_1=\kappa_2\equiv\kappa$. In a frame co-rotating with
$\omega_{\rm c}$ this solution is equivalent to a plane pendulum
where $J_z=S=0$ and was found previously~\cite{Dasgupta:2009mg}.
Shifting $\omega$ and $\gamma_{1,2}$ by $\omega_{\rm c}$ reproduces
the pure nutation solution
\begin{equation}\label{eq:solution}
{\bf P}_{\omega+\omega_{\rm c}}=
\frac{g_{\omega+\omega_{\rm c}}}
{\sqrt{\omega^4+2\omega^2\lambda^2c_{\rm m}+\lambda^4}}
\begin{pmatrix}\lambda^2 s\cr
\omega\lambda\sqrt{2(c_{\rm m}-c)}\cr
\omega^2+\lambda^2 c\cr\end{pmatrix}\,,
\end{equation}
where $s=\sin\theta$, $c=\cos\theta$, and $c_{\rm m}=
\cos\theta_{\rm min}$. In contrast with Ref.~\cite{Dasgupta:2009mg}
we here count the zenith angle $\theta$ from the north pole. In the
co-rotating frame we have $\gamma_{1,2}=\pm\gamma$ and therefore the
natural pendulum frequency is $\lambda^2=\kappa^2+\gamma^2$, whereas
the plane pendulum's highest point is given by $c_{\rm
m}=(\kappa^2-\gamma^2)/(\kappa^2+\gamma^2)$.

If the system has been set up with all polarization vectors
initially almost aligned with ${\bf B}$, the initial swap factor is
unity. The initial swap parameters are
$\gamma_1=\gamma_2=\omega_{\rm c}$ and $\kappa_1=\kappa_2=\kappa$
and are solutions of~\cite{Dasgupta:2009mg}
\begin{eqnarray}\label{eq:stability}
0&=&\int d\omega\,g_\omega\,
\frac{1}
{(\omega-\omega_{\rm c})^2+\kappa^2}\,,
\nonumber\\*
\frac{1}{\mu}&=&\int d\omega\,g_\omega\,
\frac{\omega-\omega_{\rm c}}
{(\omega-\omega_{\rm c})^2+\kappa^2}\,.
\end{eqnarray}
For the example of Fig.~\ref{fig:swapexample1} with the initial
value $\mu=10$ one finds $\omega_{\rm c}=-1.604$ and $\kappa=2.179$.

\subsection{Two pure precessions:
\boldmath$\kappa_{1,2}\ll|\gamma_2-\gamma_1|$}

When we begin with a bimodal system and the density decreases
adiabatically to zero, the end state shows a spectral swap with two
splits at its edges (Fig.~\ref{fig:swapexample1}). When $\mu$ has
become sufficiently small, the two quasi-step functions are well
separated and the overall swap factor looks like the product of two
pure precessions. This observation was the very motivation for
writing the general bimodal solution in terms of the swap parameters
$\gamma_{1,2}$ and $\kappa_{1,2}$ instead of the pendulum parameters
$\lambda$, $E$, $J_z$ and $S$.

In the present limit of $\kappa_{1,2}\ll|\gamma_2-\gamma_1|$, the
two split regions are essentially independent. The carrier modes
${\bf J}_{1,2}(t)$ precess essentially freely with frequencies
$\gamma_{1,2}$. To study this case it is easiest to represent the
$x$--$y$--components of all polarization vectors as complex numbers
of the form $J_x+\I J_y$, so the transverse components of the
carrier modes can be chosen as $\kappa_1 e^{\I\gamma_1 t}$ and
$\kappa_2 e^{\I\gamma_2 t}$ with $\kappa_{1}=\mu|{\bf J}_{1,\perp}|$
and $\kappa_{2}=\mu|{\bf J}_{2,\perp}|$. The abstract angle
characterizing the overall motion is here
$\vartheta=\pm(\gamma_1-\gamma_2)t$ because in the present limit
$\dot\vartheta^2=(\gamma_1-\gamma_2)^2$.

In the lower panels of Fig.~\ref{fig:swapexample1} we see that the
swap factor is asymmetric and steeper in the region where the step
falls into a spectral region where $g_\omega$ is smaller. This is
explained by Eq.~(\ref{eq:kappasteep}) where we have shown that for
a pure precession with very small $\mu$ the width of the step is
exponentially smaller for smaller $g_\omega$ in the step region.

As $\mu$ decreases adiabatically in a case like
Fig.~\ref{fig:swapexample1}, the system begins in a state of pure
nutation and ends in a state of two independent pure precessions.
These limiting forms and all intermediate cases are described by our
four-parameter analytic solution.

\subsection{Numerical examples}

\begin{figure}
\includegraphics[width=0.75\columnwidth]{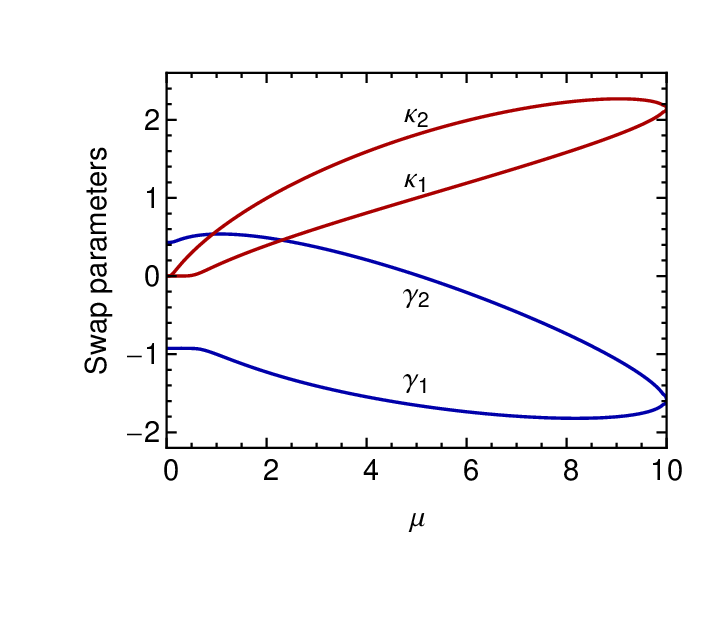}
\caption{Numerically determined swap parameters $\gamma_{1,2}$
and $\kappa_{1,2}$
for the example of Fig.~\ref{fig:swapexample1} as a function of
$\mu$ that adiabatically decreases from 10 to~0.\label{fig:adiabaticparameters}}
\end{figure}

To test if our analytic solution indeed corresponds to numerical
examples such as Fig.~\ref{fig:swapexample1}, we extract the
parameters $\gamma_{1,2}$ and $\kappa_{1,2}$ as $\mu$ slowly
decreases (Fig.~\ref{fig:adiabaticparameters}). Comparing the
analytic swap factor with the numerical one yields perfect
agreement.

\begin{figure}
\includegraphics[width=0.75\columnwidth]{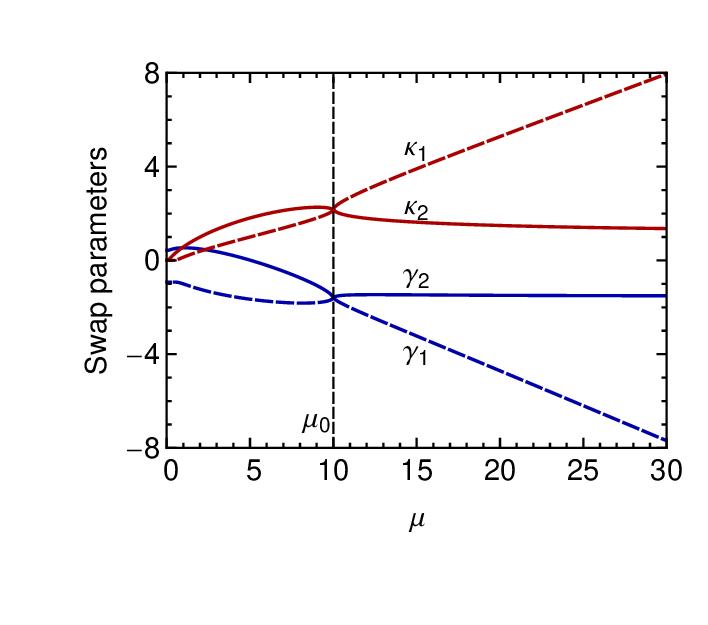}
\caption{Same as Fig.~\ref{fig:adiabaticparameters}, now with $\mu$ being
decreased from $\mu_0=10$ to zero as before, or increased from
$\mu_0=10$ to infinity.\label{fig:adiabaticparameters2}}
\end{figure}

Usually we begin with ${\bf P}_\omega^0$ almost aligned with ${\bf
B}$ and a chosen $\mu_0$. Then $\mu$ decreases from $\mu_0$ to 0,
going through different solutions as in Fig.~\ref{fig:swapexample1}
where $P_z$ remains conserved. If in this example we were to begin
with $\mu_0=3$ instead of 10, the $\mu=3$ solution would be the one
with a Lorentzian pattern. So for the same $g_\omega$, $P_z$ and
$\mu$ there exist different solutions that we can construct
numerically by adiabatic deformations.

We may also begin with a certain $\mu_0$ and then {\it increase} the
density as during supernova collapse. The initial pure nutation is
then deformed to another bimodal solution within our general
four-parameter family. For our usual example of
Fig.~\ref{fig:swapexample1} we show the swap parameters as a
function of $\mu$ in Fig.~\ref{fig:adiabaticparameters2}, assuming
the initial value is $\mu_0=10$. The solutions for $0<\mu<\mu_0$ are
from Fig.~\ref{fig:adiabaticparameters}. We identify the parameters
such that $\gamma_1<\gamma_2$.

\section{Multi-Mode Coherence}                     \label{sec:multimode}

\subsection{Four-mode coherence} \label{sec:fourmode}

\begin{figure}
\includegraphics[width=0.75\columnwidth]{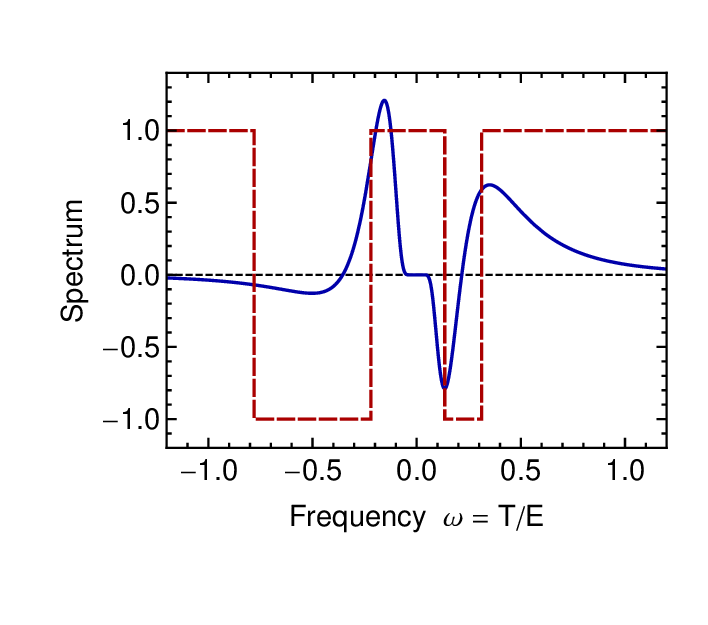}
\caption{Difference spectrum between two flavors. One is a
Fermi-Dirac distribution ($\eta=0.2$)
as in Fig.~\ref{fig:spectrum0}, the other
a nondegenerate spectrum with 1.25 larger $T$ and
0.8 times smaller $\nu$ density.
The integral over positive
$\omega$ of the degenerate flavor is normalized to unity.
Overlaid is the final swap factor after
$\mu$ has decreased adiabatically from 3 to~0.\label{fig:spectrum1}}
\end{figure}

More complicated forms of motion occur if the spectrum has two or
more independent instabilities, typically for multi-crossed
spectra~\cite{Dasgupta:2009mg}. Such spectra arise naturally if we
augment the Fermi-Dirac spectrum of a single species
(Fig.~\ref{fig:spectrum0}) with an initial population of the other
species, but with a larger temperature and smaller flux. This mimics
neutrinos streaming from a supernova core where a larger flux of
$\nu_e$ than $\bar\nu_e$ stream away and the other flavors $\nu_\mu$
and $\nu_\tau$ have smaller fluxes, larger average energies, and no
asymmetry. So if we add to the asymmetric spectrum of
Fig.~\ref{fig:spectrum0} a symmetric component of the other flavor
with a 1.25 times larger $T$ and 0.8 times smaller number density,
the difference spectrum relevant for flavor oscillations is shown in
Fig.~\ref{fig:spectrum1}.

This spectrum has two zero crossings with positive slope, allowing
for two independent instabilities, i.e.\ Eq.~(\ref{eq:stability})
can have two solutions~\cite{Dasgupta:2009mg}. If $\mu$ decreases
adiabatically from some value $\mu_0$ to 0, one finds two spectral
swaps and four concomitant splits. As an example we show the final
swap factor for $\mu_0=3$ in Fig.~\ref{fig:spectrum1}.

This case is an example for four-mode coherence. The complicated
motion of all ${\bf P}_\omega$ is equivalent to four carrier modes
as confirmed  by numerical tests using the Gram matrix. When $\mu$
has become very small and the swaps are almost complete, we have the
now-familiar pattern of four independent pure precessions in the
four split regions. In the final small-$\mu$ limit, the swap factor
is
\begin{equation}
z_\omega=\prod_{i=1}^4\,
\frac{\omega-\gamma_i}{\sqrt{(\omega-\gamma_i)^2+\kappa_i^2}}\,,
\end{equation}
corresponding to four freely precessing carrier modes.

Equation~(\ref{eq:Jhilbert2}) implies that $\bar{\bf J}_\omega^2$ is
now a polynomial with leading term $\omega^8$, so the above
representation for $z_\omega$ is general, except that in the
numerator complicated time-dependent terms appear when we are not in
the small-$\mu$ limit. One can group the four carrier modes in pairs
and understand the overall motion as two gyroscopic pendulums
interacting by a dipole force. Apart from an overall precession, the
time dependence is then described by two nutation angles and one
relative precession angle.

If we choose some value for $\mu_0$, it is not assured that the
system indeed has two instabilities---it can have only one or none.
Therefore, as $\mu$ adiabatically decreases, the system can at first
show bimodal coherence and at some critical $\mu$-value the second
unstable mode kicks in, taking the system to a state of four-mode
coherence.

\subsection{Stability issues}

It is generic that a system of lower modality can develop higher
modality by an instability. Usually we begin with all polarization
vectors almost aligned with ${\bf B}$. If they were perfectly
aligned, we would have zero-mode coherence for any spectrum
$g_\omega$, but the smallest disturbance allows the unstable modes
to grow exponentially. We implement this disturbance in the form of
a small mixing angle, i.e.\ a small mismatch between the initial
condition and exact zero-mode coherence. The solutions ($\omega_{\rm
c},\kappa$) of Eq.~(\ref{eq:stability}) identify the unstable modes,
each one corresponding to a contribution of 2 to the total modality
$N$ relevant for the given $g_\omega$ and $\mu$.

We can also prepare a state of pure precession. With an arbitrary
spectrum $g_\omega$ and parameters $\gamma$ and $\kappa$, the
initial condition is defined by
Eq.~(\ref{eq:pureprecessionsolution}), allowing us to calculate the
required $\mu$. Henceforth the system evolves in single-mode
coherent motion unless it is unstable. Depending on $g_\omega$ and
with the slightest disturbance it can transit, for example, to
three-mode coherence. Likewise, we can set up the system in some
state of bimodal coherence with suitably chosen parameters
$\kappa_{1,2}$ and $\gamma_{1,2}$, yet it can transit to higher-mode
coherence.

A formal stability criterion is only available for zero-mode
coherence in the form of Eq.~(\ref{eq:stability}) that allows us to
decide, without solving the EoMs, if the system is stuck in its
alignment with ${\bf B}$ or not.

\section{Conclusions}                          \label{sec:conclusions}

We have studied two-flavor neutrino oscillations in a homogeneous
and isotropic neutrino gas under the influence of neutrino-neutrino
refraction. This system is equivalent to an ensemble of classical
spins, each labeled by its vacuum precession frequency $\omega$
around an external magnetic field ${\bf B}$ and a dipole-dipole
interaction with identical strength $\mu$ between any pair of spins.

We have argued that the conspicuous ``self-maintained coherence''
found in this system can be understood in terms of an equivalent
``carrier system'' of a few discrete modes with the same dynamics.
We have provided an explicit construction of how the original system
depends linearly on the carrier system, assuming certain matching
conditions. We have used this approach to construct the most general
bimodal solution.

Self-maintained coherence is to be contrasted with kinematical
decoherence. We have only studied ``purely coherent'' forms of
motion, but of course, depending on initial conditions, the system
can partly or fully decohere. We have not pursued the fascinating
question of order vs.\ disorder in our
system~\cite{Pantaleone:1998xi,Raffelt:2010za}.

Our results pertain to the simplest possible toy model and as such
do not have any immediate practical impact on issues of supernova
neutrino oscillations. The next step should be to apply these ideas
to more realistic situations, notably ``multi-angle cases,'' where
isotropy is no longer assumed. The main unresolved issues in the
theory of collective neutrino oscillations in the context of
supernova neutrinos remains the role of multi-angle effects. Every
neutrino mode depends on energy and on its direction of motion,
introducing much greater complications. It would be surprising if it
were not possible to develop a more analytical understanding of
collective flavor oscillations than has been achieved by
deconstructing numerical examples. We imagine that our results are
only a first step toward a more complete theory of collective flavor
oscillations. The ultimate goal is to provide a thorough theoretical
underpinning for what one is observing in numerical simulations, and
perhaps eventually in the neutrino signal of the next galactic
supernova.

\section*{Acknowledgements} 

I thank B.~Dasgupta, A.~Dighe, A.~Patwardhan and A.~Smirnov for
discussions and critical questions and I.~Tamborra for comments on
the manuscript. This work was partly supported by the Deutsche
Forschungsgemeinschaft under grants TR-27 and EXC-153.

\section*{Note Added} 

As this paper went to press, a preprint
appeared~\cite{Pehlivan:2011hp} that studies in detail the
properties of the quantum version of our Hamiltonian,
Eq.~(\ref{eq:classHam}), and in particular its invariants. The
constants of the motion discussed in our
Appendix~\ref{sec:transformedvectors} are identified as the
so-called Gaudin magnet Hamiltonians.

\appendix

\section{Two polarization vectors}                 \label{sec:twovecs}

We briefly review the equivalence between the dynamics of two
interacting polarization vectors with that of a gyroscopic
pendulum~\cite{Hannestad:2006nj}. Consider the classical Hamiltonian
for two spin angular momenta ${\bf P}_1$ and ${\bf P}_2$ interacting
with an external magnetic field and with each other by a dipole
interaction of strength $\mu$,
\begin{equation}\label{eq:ham1}
H={\bf B}\cdot(\omega_1{\bf P}_1+\omega_2{\bf P}_2)
 +\mu\,{\bf P}_1\cdot{\bf P}_2\,.
\end{equation}
As usual, ${\bf B}$ is a unit vector in the $z$-direction and
$\omega_{1,2}$ the precession frequencies in the absence of $\mu$.
Each of ${\bf P}_{1,2}$ obeys Poisson brackets which for an angular
momentum ${\bf L}$ are $[L_i,L_j]=\epsilon_{ijk}L_k$. The Poisson
brackets of the components of ${\bf P}_1$ with those of ${\bf P}_2$
vanish and thus ${\bf P}_{1,2}^2$ is conserved. Therefore, we may
add $(\mu/2)\,({\bf P}_1^2+{\bf P}_2^2)$ to the Hamiltonian and find
\begin{equation}
H={\bf B}\cdot(\omega_1{\bf P}_1+\omega_2{\bf P}_2)
 +\frac{\mu}{2}\,{\bf P}^2\,,
\end{equation}
where ${\bf P}={\bf P}_1+{\bf P}_2$ is the total angular momentum.
The EoMs $\dot{\bf P}_{1,2}=[{\bf P}_{1,2},H]$ finally are
\begin{equation}
\dot{\bf P}_{1,2}=(\omega_{1,2}{\bf B}+\mu\,{\bf P})\times{\bf P}_{1,2}
\end{equation}
and thus the usual precession equations.

To simplify the EoMs we introduce
$\beta=\frac{1}{2}\,(\omega_1-\omega_2)$ and $\omega_{\rm
c}=\frac{1}{2}\,(\omega_1+\omega_2)$, leading to $\dot{\bf
P}_{1,2}=[(\omega_{\rm c}\pm\beta)\,{\bf B}+\mu\,{\bf P}]\times{\bf
P}_{1,2}$. The only effect of $\omega_{\rm c}$ is a common rotation
around ${\bf B}$. Therefore, we go to a rotating frame where
$\omega_{\rm c}=0$ and thus effectively $\omega_{1,2}=\pm\beta$.
Adding and subtracting these equations provides
\begin{equation}\label{eq:EoM3}
\dot{\bf Q}=\mu\,{\bf P}\times{\bf Q}
\quad\hbox{and}\quad
\dot{\bf P}=\beta\,{\bf B}\times{\bf Q}\,,
\end{equation}
where ${\bf Q}={\bf P}_1-{\bf P_2}-(\beta/\mu)\,{\bf B}$ has
conserved length. Up to a constant, the Hamiltonian becomes
\begin{equation}\label{eq:ham3}
H=\beta\,{\bf B}\cdot{\bf Q}+\frac{\mu}{2}\,{\bf P}^2\,.
\end{equation}
It is reminiscent of a gyroscopic pendulum with moment of inertia
$I=\mu^{-1}$. The first term is the ``gravitational potential'' of
the center of mass which is constrained to move on a sphere with
radius $|{\bf Q}|$. The second is the kinetic energy of the total
angular momentum ${\bf P}$.

However, the algebraic properties of ${\bf Q}$ change. Earlier its
Poisson brackets derived from its constituent angular momenta.
However, the radius-vector interpretation requires $[Q_i,Q_j]=0$ and
$[P_i,Q_j]=\epsilon_{ijk}Q_k$, where $P_i$ are the components of
${\bf P}$. Remarkably, the EoMs of Eq.~(\ref{eq:EoM3}) also follow
with the new Poisson brackets for ${\bf Q}$.

In both cases, the conserved quantities are $H$ (energy), ${\bf
B}\cdot{\bf P}$ (angular momentum along the force direction), ${\bf
Q}^2$ (length of the pendulum), and ${\bf Q}\cdot{\bf P}$ (total
angular momentum projected on the radius vector or spin). The
quantity $\lambda$ defined by
\begin{equation}\label{eq:kappadef}
\lambda^2=\beta\mu\,|{\bf Q}|
\end{equation}
plays the role of the natural pendulum frequency.

\section{Gyroscopic Pendulum}                 \label{sec:gyropendulum}

We briefly review the textbook treatment of the symmetric heavy top,
also known as Lagrangian top, gyroscopic pendulum, or spherical
pendulum with spin. It is an axially symmetric body, spinning around
its symmetry axis (moment of inertia $I_3$) and supported on this
axis. Its moment of inertia relative to the point of support is
$I_1$, mass $M$, gravitational acceleration $g$ along the
\hbox{$z$--direction}, distance $\ell$ between point of support and
center of mass, and angle $\theta$ relative to the $z$--direction.
Therefore, $V=M g\,\ell\,\cos\theta$ is the potential energy.

The angular momentum ${\bf S}$ along the symmetry axis (spin) has
kinetic energy $T_{\rm spin}={\bf S}^2/(2I_3)=I_3 \omega_{\rm
spin}^2/2$. The point of support being on the symmetry axis prevents
a torque to change $S=|{\bf S}|$ and so both $S$ and $\omega_{\rm
spin}$ are conserved. The overall motion is independent of the
internal spin angle and all tops with the same $S$ but different
$I_3$ and $\omega_{\rm spin}$ move in the same way. Therefore, we
may use a single moment of inertia $I\equiv I_1=I_3$ to describe the
system.

The orbital angular momentum is ${\bf L}=I\,{\bf q}\times{\dot{\bf
q}}$, where ${\bf q}$ is a unit vector along the symmetry axis. It
marks the top's orientation with zenith angle $\theta$ and azimuthal
angle~$\varphi$. The orbital kinetic energy is
\begin{equation}
T_{\rm orbital}=\frac{{\bf L}^2}{2I}=
\frac{1}{2}\,I\,\dot{\bf q}^2 =\frac{1}{2}\,I\,
\bigl(\dot\theta^2+\dot\varphi^2\sin^2\theta\bigr)\,.
\end{equation}
The total angular momentum ${\bf J}={\bf L}+{\bf S}$ has conserved
\hbox{$z$--component}, where $S_z=S\,\cos\theta$ and
$L_z=I\dot\varphi\sin^2\theta$. Here one factor of $\sin\theta$
comes from the projection of ${\bf q}$ on the transverse plane and
the velocity is $\dot\varphi\sin\theta$. Therefore,
$J_z=I\dot\varphi\sin^2\theta+S\cos\theta$ is conserved and
\begin{equation}\label{eq:phimotion}
\dot\varphi=\frac{J_z-S\cos\theta}{I\sin^2\theta}\,.
\end{equation}
Therefore
\begin{equation}
T_{\rm orbital}=\frac{1}{2}\,I\,\dot\theta^2+
\frac{(J_z-S\cos\theta)^2}{2 I\sin^2\theta}
\end{equation}
and the total energy $E=T+V$ is
\begin{equation}
E=\frac{I}{2}\,\dot\theta^2+
\frac{(J_z-S\cos\theta)^2}{2 I\sin^2\theta}
+M g \ell \cos\theta\,.
\end{equation}
This has the form $E=I\,\dot\theta^2/2+V(\theta)$, where $V(\theta)$
is a potential given in terms of conserved quantities fixed by
initial conditions. One may now solve for $\theta(t)$ by a
quadrature and then find $\varphi(t)$ by integrating
Eq.~(\ref{eq:phimotion}).

Next we introduce $c=\cos\theta$ as independent variable so that
$\dot\theta^2=\dot c^2/\sin^2\theta$ and find the usual third-order
polynomial in $c$,
\begin{equation}\label{eq:thetadot3}
\dot c^2=2\,\frac{E-Mg\ell\,c}{I}\,(1-c^2)-\frac{(J_z-S\,c)^2}{I^2}\,,
\end{equation}
that can be integrated using elliptic functions. A typical motion is
nutation between two limiting angles $\theta_1$ and $\theta_2$ as
well as precession. When $J_z=S=0$ we have a plane pendulum where
\begin{equation}
\lambda^2=M g\,\ell/I
\end{equation}
gives us the natural frequency $\lambda$.

The motion of two interacting polarization vectors is equivalent to
a gyroscopic pendulum (Appendix~\ref{sec:twovecs}) and therefore can
be explicitly solved in terms of elliptic functions. This feat was
achieved previously without invoking the equivalence to a
pendulum~\cite{Samuel:1996ri}.

\section{Transformed ensemble of polarization vectors}
\label{sec:transformedvectors}

The functions $\bar{\bf J}_\omega(t)$ derived from the carrier modes
by Eqs.~(\ref{eq:linearcombi2}) or~(\ref{eq:Jhilbert2}) represent a
special case of a more general transformation of an ensemble ${\bf
P}_\omega(t)$ obeying Eq.~(\ref{eq:precession}). To motivate this
transformation we observe that in the non-interacting case, every
${\bf B}\cdot{\bf P}_\omega$ is conserved. Is there a similar
conserved quantity in the interacting case?

So we look for a vector ${\bf X}$ such that ${\bf X}\cdot{\bf
P}_\omega$ is conserved. A trivial example is ${\bf X}={\bf
P}_\omega$, but any other ${\bf X}$ fulfilling $\dot {\bf X}=(\omega
{\bf B}+\mu {\bf P})\times{\bf X}$ is a solution. To see this
consider $\dot{\bf X}(t) = {\bf V}(t)\times{\bf X}(t)$, where ${\bf
V}(t)$ is externally prescribed and solve for ${\bf X}(t)$ with
initial condition ${\bf X}_0$. Another initial condition ${\bf Y}_0$
provides ${\bf Y}(t)$ and ${\bf X}(t)\cdot{\bf Y}(t)$ is conserved.
Consider $d({\bf X}\cdot{\bf Y})/dt=\dot{\bf X}\cdot{\bf Y}+{\bf
X}\cdot{\dot{\bf Y}} =({\bf V}\times{\bf X})\cdot{\bf Y} +{\bf
X}\cdot({\bf V}\times{\bf Y})$ and after a permutation in one of the
triple products one finds $d({\bf X}\cdot{\bf Y})/dt=0$.

We seek ${\bf X}$ as a linear combination of the ${\bf P}_\omega$
and also of ${\bf B}$ because ${\bf X}={\bf B}$ is a solution for
$\mu\to0$. The length of ${\bf X}$ is arbitrary, so we assume the
form ${\bf X}={\bf B}+\int d\omega'\,a_{\omega'}{\bf P}_{\omega'}$
and seek the coefficients~$a_{\omega'}$. The requirement $\dot {\bf
X}=(\omega {\bf B}+\mu {\bf P})\times{\bf X}$ implies $\int
d\omega'\,a_{\omega'}\dot{\bf P}_{\omega'}= (\omega {\bf B}+\mu {\bf
P})\times\left({\bf B}+ \int d\omega'\,a_{\omega'}{\bf
P}_{\omega'}\right)$. On the l.h.s.\ we insert $\dot{\bf
P}_{\omega'}=(\omega'{\bf B}+\mu{\bf P})\times{\bf P}_{\omega'}$,
whereas on the r.h.s.\ we use ${\bf B}\times{\bf P}={\bf
B}\times\int d\omega'{\bf P}_{\omega'}$. Collecting all terms we
find $0={\bf B}\times\int d\omega'\,
\left[(\omega-\omega')a_{\omega'}-\mu\right]\,{\bf P}_{\omega'}$ and
thus $a_{\omega'}=\mu/(\omega-\omega')$.

We are therefore led to define the linearly transformed polarization
vectors
\begin{equation}\label{eq:hilbert}
{\bar{\bf P}}_\omega(t)={\bf B}+
\mu\int d\omega'\,
\frac{{\bf P}_{\omega'}(t)}{\omega-\omega'}\,.
\end{equation}
The integral is essentially the Hilbert transform in the $\omega$
variable and is understood as a Cauchy principal value. (The Hilbert
transform of a function $f(x)$ is defined as its convolution with
$1/x$.) So we have found a unique linear combination $\bar{\bf
P}_\omega(t)$ of the original ensemble ${\bf P}_\omega(t)$ that
obeys the original precession equation in the form
\begin{equation}
\partial_t{\bar{\bf P}}_\omega=(\omega{\bf B}+\mu{\bf
P})\times\bar{\bf P}_\omega\,.
\end{equation}
As a consequence, ${\bf P}_\omega^2$, $\bar{\bf P}_\omega^2$, and
${\bf P}_\omega\cdot\bar{\bf P}_\omega$ are conserved for any
fixed~$\mu$.

The intriguing properties of the transformed ensemble remain to be
explored. We believe, for example, that coherent oscillations
correspond to ${\bf P}_\omega(t)$ and $\bar{\bf P}_\omega(t)$ being
collinear at all times.


\end{document}